\documentclass{ethpaper}
\usepackage{graphicx}
\usepackage{latexsym,rotating,subfigure}
\usepackage{subfigure}
\usepackage{wrapfig,epsfig}

\begin{document}

\begin{titlepage}

   \ethnote{}
\title{Crystals for high-energy calorimetry in extreme environments}

\begin{Authlist}
F.~Nessi-Tedaldi
\Instfoot{eth}{Swiss Federal Institute of Technology (ETH),
CH-8093 Z\"urich, Switzerland}
  \end{Authlist}
\maketitle
\abstract{Crystals are used as a homogeneous calorimetric medium in many
high-energy physics experiments. For some experiments, performance has
to be ensured in very difficult operating conditions, like a high
radiation environment, very large particle fluxes, high collision
rates, placing constraints on response and readout time.  An overview
is presented of recent achievements in the field, with particular
attention given to the performance of Lead Tungstate (PWO) crystals
exposed to high particle fluxes.}
\vspace{7cm}
\conference{Presented at the
{Meeting of the Division of Particles and Fields of the
American Physical Society\\DPF2004}\\ Riverside, USA,
August 26th to 31st, 2004}

\end{titlepage}

\setcounter{page}{2}
This report is a brief overview on the existing knowledge about
scintillating crystals performance, when they are used for high-energy
physics calorimetry, with particular attention given to the behavior
under high radiation levels and intense particle fluxes.
The way high ionizing radiation levels
affect crystal performance will only be briefly summarized, since it
has been thoroughly studied and reported upon by many authors
in the past, as crystals were used e.g. in $e^+e^-$ collider experiments,
and their growth
parameters were optimized for best performance in such environments.
Today's renewed concern deals with crystals exposed to high
particle fluxes, running conditions expected in several
experiments under construction or designed, such as hadron collider
detectors. Some recent and older results are thus collected herein, to
provide, as far as possible, a complete picture.

Ionizing radiation produces absorption bands through formation of
color centers, which reduce the Light Transmission (LT) and thus the
Light Output (LO), whose importance is related to oxygen contamination
in alkali halides like $\mathrm{BaF}_2$ and CsI, and to oxygen
vacancies and impurities in oxides like BGO and
PWO~\cite{r-ZHU1}. Phosphorence or afterglow may also
appear~\cite{r-ETH1}, which increase the noise levels in the detected
light, possibly worsening the energy resolution (in a negligible way
for PWO in LHC experiments~\cite{r-ZHU2}), while the scintillation
mechanism is not damaged. Recovery of damage at room temperature can
occur depending on crystal type, and growth parameters, giving rise to
a dose-rate dependence of damage equilibrium levels~\cite{r-ETH2}, and
to a recovery speed dependent on the depth of traps.  That ionising
radiation only affects LT, means the damage can be monitored through
light injection and corrected for, as it is done in the CMS
Electromagnetic Calorimeter (ECAL)~\cite{r-Bornheim}.

The way hadron fluxes affect crystals has been studied to a lesser
extent, but crucial questions arise while detectors are being
constructed, namely, whether such fluxes cause a specific, possibly
cumulative damage, and if so, what its quantitative importance is,
whether it only affects LT or also the scintillation mechanism. Tests
were recently performed on PWO at IHEP Protvino~\cite{r-Batarin} for
the BTeV experiment and at CERN and ETH-Z\"urich~\cite{r-ETH3} for the
CMS ECAL.

In BTeV, 94\% of the crystals will see relatively low particle fluxes,
corresponding to ionizing dose rates below 36 rad/h and doses up to
100 krad/y, while at the other end, $0.5$~\% of the crystals will
experience high fluxes ($\sim200$ rad/h and beyond, up to 2 Mrad/y). These
regimes were tested in crystals using $e^-$ and $\pi$ beams and
$\gamma$ sources up to a few krad at 1 to 60 rad/h at one end, and a very
intense mixed beam of charged hadrons, neutrons and $\gamma$ up to
3 Mrad at 1 krad/h and 100 krad/h equivalent fluxes at the other
end. In $e^-$, $\pi$ and $\gamma$ irradiations, the signal loss
behavior is qualitatively similar between electrons and pions, and the
damage appears to reach equilibrium at a dose-rate dependent
level. Furthermore, no indication of damage to the scintillation mechanism
from $\pi$ irradiation is found~\cite{r-Batarin3}. A concern remains
however, that for the dose-rates used, the total absorbed dose
expected in the experiment is not explored and thus an additional,
specific, possibly cumulative damage from hadrons cannot be
excluded. This concern is partially answered by the irradiations in
the very intense, mixed beam, where the absorbed doses correspond to 3
y running for the most exposed crystals, yielding thus an upper limit
to the damage expected in the experiment. At the constant flux used,
the damage appears in fact to be steadily increasing with accumulated
dose. This is unlike pure ionizing radiation damage, which reaches
equilibrium at a level depending on dose rate, not beyond what
saturation of all color centers can yield, and thus it constitutes an
indication for an additional, cumulative, hadron-specific damage.

For CMS, hadron fluences have been calculated~\cite{r-ETH3} for
$5\times 10^5\;\; \mathrm{pb}^{-1}$ (10 y running at LHC), yielding in
the ECAL barrel (end caps) $\sim 10^{12}\; (\sim 10^{14})\;
\mathrm{cm}^{-2}$ .  Hadron-specific damage could be due to the
production, above a $\sim 20$ MeV threshold, of heavy fragments
(``stars''), with up to 10 $\mu$m range and energies up to $\sim$100
MeV, which cause displacement of lattice atoms and energy losses along
their path up to 10000 $\times$ the one of minimum-ionizing
particles. The damage caused by these processes is likely different
from the one of ionizing radiation, thus possibly cumulative. The
primarily investigated quantity was the induced absorption coefficient
at peak-of-emission wavelength, defined as
$\mu_{IND}\mathrm{(440\; nm)}=\frac{1}{L}\ln\frac{LT0}{LT}$, with
LT0 and LT the longitudinal
transmission at 440 nm before and after irradiation, respectively,
measured through the crystal length L. Transmission measurements are
very accurate and directly related to LO changes provided
scintillation is not affected. For CMS production crystals,
$\mu_{IND}\mathrm{(440 nm)}= 1\;\mathrm{m}^{-1}$ corresponds to a
relative LO loss of 25\%~\cite{r-AUFFRAY}.

Several CMS production crystals of consistent quality\footnote{Prime
  indicates a second irradiation of the same crystal.}
  were irradiated in a 20 GeV proton flux of
$10^{12}\;\mathrm{p/cm}^{2}/\mathrm{h}$ ({\em a,b,c,d}) and a few more
({\em E, F, G}) in $10^{13}\;\mathrm{p/cm}^{2}/\mathrm{h}$.
\label{s-CMS}
\begin{figure}[h]
\begin{center}\footnotesize
  \begin{tabular}[h]{cc}
 \subfigure[LT for crystals {\em a} and {\em F} \label{f-LT} before,
after a first and after a second irradiation.] 
{\mbox{\includegraphics[width=65mm]{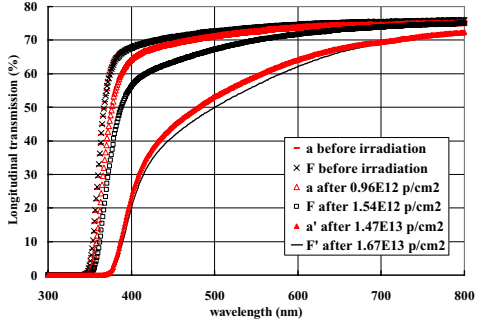}}}&
 \subfigure[Correlation between $\mu_{IND}\mathrm{(440 nm)}$
and proton fluence.\label{f-mu}]
{\mbox{\includegraphics[width=67mm]{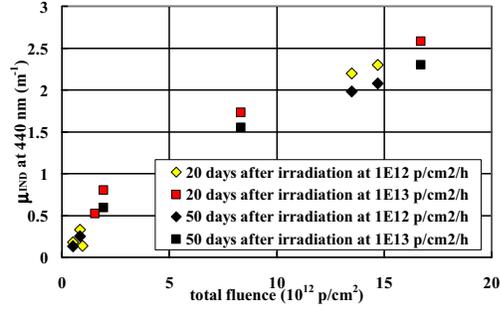}}}
   \end{tabular}
\end{center}
\caption{Proton irradiation results of PWO\label{f-PWO}}
\end{figure}
The LT data in Fig.~\ref{f-LT} show a possible flux dependence at low
fluence, no flux dependence at high fluence and a transmission
band-edge shift, unlike what happens in purely ionizing
radiation~\cite{r-ZHU1}.
The correlation in Fig.~\ref{f-mu}, between $\mu_{IND}\mathrm{(440 nm)}$
and fluence, is consistent with a linear behavior, showing
that proton-induced damage in PWO appears to predominantly yield a
cumulative effect. Taking into account the difference in composition
and energy spectra between 20 GeV protons and CMS, the test results cover
\begin{wrapfigure}{r}{70mm}
\mbox{\epsfig{file=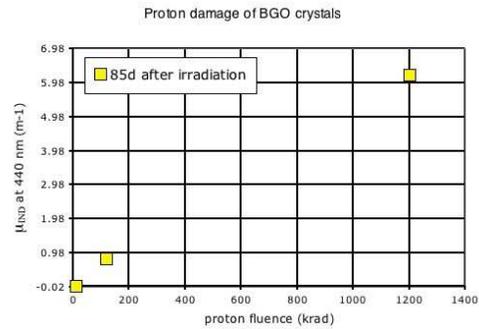,width=65mm}}
\caption{Correlation between $\mu_{IND}\mathrm{(440 nm)}$ and proton
fluence in BGO extracted from published data.\label{f-BGO}}
\end{wrapfigure}
the CMS running conditions over the entire high-precision ECAL
pseudo-rapidity region, up to $\sim$~$2.9$.

Proton and $\gamma$ data are compared in a study performed on
BGO~\cite{r-KobaBGO}. The changes in band-edge are similar to what is
seen in PWO, and long enough after irradiation, when the
ionising-radiation damage contribution has recovered, one can extract
a remaining p-damage that behaves linearly with fluence, as
visible in Fig.~\ref{f-BGO}.  The same exercise is not possible on CsI
data from the same authors~\cite{r-KobaCsI} because the damage caused
by ionizing radiation gives a too important constribution to allow
observing the proton-specific one.

In conclusion, one can say that for all crystals commonly used in
calorimetry, beyond the well-studied damage from ionizing radiation,
the understanding of additional contributions to the damage, when
crystals experience a substantial hadron flux, has become important
since experiments are being built having to cope with such running
conditions. A hadron-specific, cumulative contribution, likely due to
the intense local energy deposition from heavy fragments, has been
observed in tests on PWO and BGO.  Within the explored flux and
fluence ranges, this contribution only seems to affect Light
Transmission, and thus can be monitored. Additional studies, taking
into account the range of operation of the envisaged experiments, are
expected to consolidate the present understanding of hadron damage.

\end{document}